\documentstyle[12pt]{article}   
\def\e{\mbox{e}}                
\def\mathrm#1{#1}               
\newenvironment{ack}{\subsection*{Acknowledgenements}}{}  
\font\twlmib = cmmib10 scaled \magstep1
\def\vec#1{\mbox{\twlmib #1}}
\begin{document}
\title{The Delta-Hole Model\\ at Finite
       Temperature\thanks{Work supported by GSI}}
\author{P.A.Henning\thanks{E-mail address: phenning@tpri6a.gsi.de}\\
        Institut f\"ur Kernphysik der TH Darmstadt and GSI\\[1mm]
        P.O.Box 110552, D-6100 Darmstadt, Germany\\[3mm]
        and\\[3mm]
        H.Umezawa\thanks{E-mail address: umwa@phys.ualberta.ca}\\
        The Theoretical Physics Institute,
        University of Alberta\\[1mm]
        Edmonton, Alberta T6G 2J1, Canada}
\date{1 April, 1993}
\begin{abstract}
The spectral function of pions interacting with a
gas of nucleons and $\Delta_{33}$-resonances is investigated
using the formalism of Thermo Field Dynamics. After a
discussion of the zero $\Delta$-width approximation at finite
temperature, we take into account a constant width of the resonance.
Apart from a full numerical calculation, we give analytical
approximations to the pionic spectral function including such a width.
They are found to be different from previous approximations,
and require an increase of the effective $\Delta$-width in hot
compressed nuclear matter. The results are summarized in an
effective dispersion relation for interacting pions.
\end{abstract}
\maketitle           
\clearpage
\section{Introduction}
One of the main {\em experimental\/} goals of modern nuclear physics
is the investigation of hot, compressed nuclear matter
(HCNM).  {\em Theoretical\/} descriptions of HCNM are mostly
based on, or at least motivated by, methods of relativistic
quantum field theory. However, within this framework it is a well
established fact, that naive perturbation theory breaks down at
finite temperature \cite{K89,E90}.  This is due the modification of
space-time symmetry in the presence of matter or a heat bath,
i.e.  to the absence of stable asymptotic states for the
observable physical particles \cite{L88}.
To overcome this problem, first of all one has to
use one of several existing
formulations of quantum field theory with $2\times 2$
matrix-valued propagators.

While so far the Closed-Time-Path method (CTP)
has been used to derive transport
equations for heavy-ion collisions \cite{SKM,MH93}, the model
called Thermo Field Dynamics (TFD) offers technical
advantages over CTP \cite{AU87,Ubook}. Furthermore, within such a formulation,
proper asymptotic conditions necessary for a perturbative approach
can be defined only in TFD
(see ref. \cite{HU92} for details).

In this paper, TFD is applied to a system potentially
interesting for the theoretical description of HCNM,
namely to an interacting gas of nucleons, pions
and $\Delta_{33}$-resonances \cite{WEISE}.
This $\Delta$-hole model is often used
\cite{EWC93,KXS89,XKK93}, but in our view the analytical structure
of the pion propagator is far from well understood.

We first describe the model as found in many applications, i.e.
in the quasistatic zero-$\Delta$-width approximation.
Then a more realistic Lorentz-type spectral function for the
$\Delta$'s is discussed, and the full pion propagator is
calculated from a dispersion integral.

To obtain a systematic extension
of the simple $\Delta$-hole model to finite $\Delta$-width
and finite temperature, analytical results are given
for the pion propagator by introducing an asymptotic expansion
of the $\Delta$ spectral function to first order in its width.
These results are simple enough to be used in further
applications of the model, but still complete enough to
contain the full information of the $\Delta$-hole model.

It is discussed, how to fit this simplified model to the full
calculation, and how good the approximations are at various
temperatures, densities and momenta. Finally, the model
is used to extract an ''effective'' pion dispersion relation
in HCNM.

In the framework of TFD, the thermal instability of observable
states can be absorbed into a Bogoliubov transformation also for
interacting systems. This Bogoliubov transformation defines
stable quasi-particles which serve as basis for a perturbation
expansion. It can  be written in a single matrix form for the
bosonic and the fermionic sector of the
model as
\begin{equation}\label{bdef}
 {\cal B}_{B,F}(n) =
\left(\array{cc}\left(1 \pm n\right) & -n\\
                \mp 1 & 1\endarray\right)
\;.\end{equation}
The Bogoliubov parameters $n$ are the phase-space distribution
functions for bosons or fermions.
While one has some freedom in parameterizing this transformation, the
above form was found to be the most useful in ref. \cite{Ubook},
since it makes the propagators linear in $n$.

In the fermionic sector, i.e., for nucleons and $\Delta$'s,
the $n$  are Fermi-Dirac functions
\begin{equation}
n_{N,\Delta}(E)   =   \frac{1}{\e^{\beta(E-\mu_{N,\Delta})}+1}
\;,\end{equation}
with inverse temperature $\beta$. The baryon number in each small
volume and hence the baryon density is a constant parameter of
the calculations. Forgetting about the interaction at the moment,
we express it in terms of bare ''on-shell'' energies
\begin{equation}
E_N(\vec{p})  =  \sqrt{\vec{p}^2 + M_N^2}\;\;\;\mbox{and}\;\;\;
E_\Delta(\vec{p})  =  \sqrt{\vec{p}^2 + M_\Delta^2}
\;,\end{equation}
and obtain as the baryon density without interaction
\begin{eqnarray}\label{rb0}
\rho_b^0& =& \rho^0_N + \rho^0_\Delta \nonumber \\
      & =& 4\,\int\!\!\frac{d^3\vec{p}}{(2\pi)^3}\,n_N(E_N(\vec{p}))
          +16\,\int\!\!\frac{d^3\vec{p}}{(2\pi)^3}\,n_\Delta(E_\Delta(\vec{p}))
\;.\end{eqnarray}
In the fermion propagators we neglect anti-particle states.
The resulting $2\times 2$-matrices are
\begin{eqnarray}\label{fbk}
&&S^{(ab)}_{N,\Delta}(p_0,\vec{p}) =
 -\mathrm{i}\,\int\!\!dE\,{\cal A}_{N,\Delta}(E,\vec{p})\;\times\nonumber \\
&&\tau_3\, ({\cal B}_F(n_{N,\Delta}(E)))^{-1}\;
   \left(\!{\array{ll}
         {\displaystyle \frac{1}{p_0-E+\mathrm{i}\epsilon}} & \\
    &    {\displaystyle \frac{1}{p_0-E-\mathrm{i}\epsilon}}
\endarray}\right)\;
 {\cal B}_F(n_{N,\Delta}(E))
\;\end{eqnarray}
for both nucleons and $\Delta$'s. In this equation,
${\cal A}_{N,\Delta}$ is the spectral function of nucleons or $\Delta$'s,
and ${\cal B}_F(n_{N,\Delta})$ are Bogoliubov matrices containing the
thermal information as defined above, $\tau_3=\mathrm{diag}(1,-1)$.
\section{Full pion propagator}
The full pion propagator is the solution of a
Schwinger-Dyson equation, with the free propagator and
a self energy function (polarization function) as input.
Each of these quantities is $2\times 2$-matrix valued,
and the full as well as the free propagator obey
\begin{equation}\label{lpp}
D^{11}+D^{22} = D^{12} + D^{21}
\;.\end{equation}
It is easy to see, that this implies for the polarization function
\begin{equation}\label{lp}
\Pi^{11}+\Pi^{12}+\Pi^{21}+\Pi^{22} = 0
\;.\end{equation}
In the following, we use results derived in ref. \cite{HU92}.
{}From the linear relation (\ref{lp}) between the matrix elements one
obtains, that the matrix valued polarization function can
be brought to triangular form by multiplication with
bosonic Bogoliubov matrices ${\cal B}_B$
\begin{eqnarray}\label{dp}
&{\cal B}_B(n_\pi(E))
   \,\tau_3\,\left( \vphantom{\int} \Pi^{(ab)} \right)&
   \,({\cal B}_B(n_\pi(E)))^{-1} \nonumber\\
&  =&\left( {\array{lr} \Pi^{11}+\Pi^{12} & (1+n_\pi(E))\Pi^{12}
   -n_\pi(E)\Pi^{21} \\  0 &\Pi^{11}+\Pi^{21} \endarray}\right)
\;\end{eqnarray}
for {\em any\/} value of $n_\pi(E)$.
The diagonal elements are the
retarded and advanced components of the polarization
\begin{eqnarray}\label{prd}
\Pi^R &= &\Pi^{11}+\Pi^{12}\nonumber \\
\Pi^A &= &\Pi^{11}+\Pi^{21}
\;.\end{eqnarray}
As one can derive, the off-diagonal components of $\Pi$
are related by
\begin{equation}\label{lq}
\Pi^{12}(E,\vec{k})-
  \e^{-\beta(E-\mu_\Delta+\mu_N)}\,\Pi^{21}(E,\vec{k})=0
\;.\end{equation}
Note, that this relation is quite similar (and indeed equivalent)
to the Kubo-Martin-Schwinger boundary condition for propagators,
which is equivalent to an equilibrium condition.

The full pion propagator therefore can be {\em diagonalized\/} by
Bogoliubov matrices, if and only if $n_\pi(E)$ is chosen such that
the off-diagonal element in (\ref{dp}) vanishes,
i.e. by setting
\begin{equation}
n_\pi(E) = \frac{1}{\e^{\beta(E-\mu_\Delta+\mu_N)}-1}
\;.\end{equation}
Since $\mu_\pi=\mu_\Delta-\mu_N$ in our simple model,
this is the pionic equilibrium
distribution function, and thus the pion propagator is diagonal
(separately for positive and negative energy states) only in case
the pions have the same temperature as the baryons.
For the results presented in this paper we have used
$\mu_\pi=0$, but the equations can be applied also to cases
with nonzero pion chemical potential.

The full pion propagator is
\begin{eqnarray}\label{dbk}
D^{(ab)}(k_0,\vec{k})&& = \int\!\!dE\,
 {\cal A}_\pi(E,\vec{k})\;\times\nonumber\\
  &&\left( \;({\cal B}_B(n_\pi(E)))^{-1}
  \left(\!{\array{ll}
         {\displaystyle \frac{1}{k_0-E+\mathrm{i}\epsilon}} & \\
    &    {\displaystyle \frac{1}{k_0-E-\mathrm{i}\epsilon}}
\endarray} \right)\;
         {\cal B}_B(n_\pi(E))\,\tau_3 \right.\\
  -&&\left. \tau_3\,{\cal B}^T_B(n_\pi(E))\;
  \left(\!{\array{ll}
  {\displaystyle \frac{1}{k_0+E-\mathrm{i}\epsilon}} & \\ &
  {\displaystyle \frac{1}{k_0+E+\mathrm{i}\epsilon}}
\endarray}\right)\;
          ({\cal B}^T_B(n_\pi(E)))^{-1}\right)\nonumber
\;.\end{eqnarray}
${\cal A}_\pi(E,\vec{k})$ is a positive function,
the spectral function of the pion field. Thus, apart from the task of
determining this function, (\ref{dbk}) already solves the
problem of pion propagation in HCNM.

The spectral function is related to the
imaginary part of the retarded and advanced propagator,
\begin{equation}\label{rap}
D^{R,A}(E,\vec{k})  =
  \int\!\!dE^\prime\;{\cal A}_\pi(E^\prime,\vec{k})\;
   \left(\frac{1}{E-E^\prime\pm \mathrm{i}\epsilon}
        -\frac{1}{E+E^\prime\pm \mathrm{i}\epsilon} \right)
\;,\end{equation}
and the limit of free particles is recovered when
\begin{equation}\label{fb}
{\cal A}_\pi(E,\vec{k}) \longrightarrow
  \delta(E^2-E^2_\pi(\vec{k}))\Theta(E)
\;,\end{equation}
with the bare ''on-shell'' energy
\begin{equation}
E_\pi(\vec{k}) = \sqrt{\vec{k}^2 + m^2_\pi}
\;.\end{equation}
For the interacting system, the propagator is best expressed through
retarded and advanced components of the polarization
since their analytical structure is well defined. One has
\begin{eqnarray}\label{pisd}
\Pi^{R,A}(E,\vec{k}) &=& \mbox{Re}\left(\Pi(E,\vec{k})\right)
   \mp \mathrm{i}\,\pi\sigma(E,\vec{k})\nonumber\\
& =&\int\!\!dE^\prime\;\frac{\sigma(E^\prime,\vec{k})}{
                      E-E^\prime\pm \mathrm{i}\epsilon}
\;.\end{eqnarray}
The Schwinger-Dyson equation can be solved directly using
this decomposition. In our formalism it is then diagonal, corresponding
to two complex conjugate equations for the
retarded and advanced propagator. The solution is
\begin{equation}\label{rsc}
{\cal A}_\pi(E,\vec{k}) =
   \frac{\sigma(E,\vec{k})\,\Theta(E)}{
         \left(E^2-E_\pi^2(\vec{k})-\mbox{Re}\left( \Pi(E,\vec{k})\right)
\right)^2+
         \pi^2\sigma^2(E,\vec{k})}
\;.\end{equation}
In the limit of vanishing self energy the free-field
result (\ref{fb}) is recovered.

We now use the fermion propagators (\ref{fbk}) to obtain the
polarization function for pions coupled to the fermions.
Although we make a one-loop calculation, this does by no means
imply a perturbative character of the calculation, since we
can, in principle, use the {\em full\/} Green's function
from the fermionic sector. All the interaction effects of the
baryons are then plugged in their spectral functions,
missing is only a possible vertex correction term.
Such a vertex correction term can be parameterized as a medium dependence
of the coupling constant.

In this generalized one-loop approximation, one has
\begin{eqnarray}\label{pidef}
\Pi_0^{(ab)}(E,\vec{k})\;=\;
 -\mathrm{i}\,s\,\vec{k}^2
\;\int\!\!\frac{d^4p}{(2\pi)^4}\,
  \left[g^{(a)} S^{(ab)}_N(p) g^{(b)} S^{(ba)}_\Delta(p-k)
  \;\right.&&\nonumber \\
  \left.g^{(a)} S^{(ab)}_N(p) g^{(b)} S^{(ba)}_\Delta(p+k)\right]&&
\;\end{eqnarray}
where $s=16/9$ is a spin degeneracy factor, $k=(E,\vec{k})$
in the integrand and
\begin{equation}
g^{(1)}=-g^{(2)}=\frac{f^\pi_{N\Delta}}{m_\pi}
\;.\end{equation}
The numerical values for the coupling parameters and masses
can be found in table 1.
\begin{table}[b]
\hrule
\vspace*{2mm}
\begin{center}
\begin{tabular}{cccccc} \hline
{}~$f^\pi_{N\Delta}$~ & ~$g^\prime$~ & ~$m_\pi$~ & ~$M_N$~     &
         ~$M_\Delta$~ & ~$\Gamma$~ \\ \hline
         2          & ~0.5~        & ~0.14 GeV~& ~0.938 GeV~ &
          ~1.232 GeV~ & ~0.12 GeV~  \\ \hline
\end{tabular}
\end{center}
\caption{Coupling constants and masses used in the calculations of
 this work.}
\end{table}

{}From the definition (\ref{pidef}) one can easily
derive the relations (\ref{lp}) and (\ref{lq}).
Inserting the fermion propagators given above then leads to
a retarded pionic polarization function
\begin{eqnarray}\label{rpp}
\Pi^R(E,\vec{k}) & =&  -2\pi\,\vec{k}^2\,
  \frac{16}{9}\left(\frac{f^\pi_{N\Delta}}{m_\pi}\right)^2\,
  \int\!\!\frac{d^4p}{(2\pi)^4}
  \;\int\!\!dz_1\,dz_2\,{\cal A}_N(z_1,\vec{p}-\vec{k})\,
  {\cal A}_\Delta(z_2,\vec{p})\,\times\nonumber\\
&&\left(\,\delta(p_0-E-z_1)\,
  \frac{n_N(z_1)}{p_0-z_2-\mathrm{i}\epsilon}\;
 +\;\delta(p_0-z_2)\,
  \frac{n_\Delta(z_2)}{p_0-E-z_1+\mathrm{i}\epsilon}\right)\\
&+&\mbox{the same expression with
$E\rightarrow -E$, $\epsilon\rightarrow -\epsilon$}\nonumber
\;.\end{eqnarray}
The imaginary part of this expression determines the function
$\sigma(E,\vec{k})$ as
\begin{eqnarray}
\sigma(E,\vec{k})&=&\vec{k}^2
  \frac{16}{9}\left(\frac{f^\pi_{N\Delta}}{m_\pi}\right)^2\,
  \int\!\!\frac{d^3\vec{p}}{(2\pi)^3}
  \;\int\!\!dz\,{\cal A}_N(z,\vec{p})\,\times\nonumber\\
&&\left({\cal A}_\Delta(z+E,\vec{p}+\vec{k})\vphantom{\int}
      \left(\,n_N(z)-n_\Delta(z+E)\,\right)\right.\\
&-&\;\;\;\left.{\cal A}_\Delta(z-E,\vec{p}+\vec{k})\vphantom{\int}
      \left(\,n_N(z)-n_\Delta(z-E)\,\right)\right)\nonumber
\;.\end{eqnarray}
This function rather than (\ref{rpp}) is our starting point for
the calculation of the full pion propagator, since it is
numerically much easier to calculate.  To include vertex
corrections, one could then multiply $\sigma(E,\vec{k})$ by a form
factor like in ref. \cite{XKK93}, which however we will avoid
here.  The next step consists in calculating the dispersion
integral (\ref{pisd}).

Finally, we add another piece which has not been addressed
so far. The perturbative expression for the pionic polarization tensor
in the medium has to be corrected for the strong
repulsive interaction at short distances. We use
the phenomenological description
\begin{equation}\label{efp}
\Pi_c(E,\vec{k})= \frac{\vec{k}^2\,\Pi(E,\vec{k})}{
        \vec{k}^2-g^\prime\,\Pi(E,\vec{k})}
\;.\end{equation}
Note, that in HCNM $g^\prime$ also depends on the system parameters.
\section{Quasistatic zero-width approximation}
A first approximation to the $\Delta$-hole model is obtained
with the {\em free\/} spectral functions for nucleons and $\Delta$'s,
ignoring the finite lifetime of the latter.
However, since this model is widely used, its critical discussion is
necessary in this work. Hence we set for the nucleons
\begin{equation}\label{an}
{\cal A}_N(E,\vec{p}) = \delta(E-E_N(\vec{p}))
\;,\end{equation}
and neglect the width of the $\Delta$-particle by using
\begin{equation}\label{ad}
{\cal A}_\Delta(E,\vec{p}) = \delta(E-E_\Delta(\vec{p}))
\;.\end{equation}
The baryon density of the system is then given by
the expression (\ref{rb0}).

The quasistatic approximation follows from this by expanding the
$\delta$-function in the integrand in powers of $\vec{p}/M_N$.
To lowest order, it corresponds to the neglection of recoil effects,
i.e. by setting
\begin{equation}
\delta(E_\Delta(\vec{p}+\vec{k})-E_N(\vec{p})-E)
   \approx\delta(E_\Delta(\vec{k})-M_N-E)
\;\end{equation}
in the integrand. The $\delta$-functions therefore do not affect the
momentum integration, and one obtains
\begin{eqnarray}\label{sig0}
\sigma_{q0}(E,\vec{k}) & =&\vec{k}^2
  \frac{16}{9}\left(\frac{f^\pi_{N\Delta}}{m_\pi}\right)^2\,
  \left\{\int\!\!\frac{d^3\vec{p}}{(2\pi)^3}\,
      \left(\,n_N(E_N(\vec{p}))-n_\Delta(E_\Delta(\vec{p}))\,
      \right)\right\}\,\times\nonumber\\
 &&\;\;\;\;\;\;\;\;\;
 \left( \delta(E-E_\Delta(\vec{k})+M_N)\,\vphantom{\int}
         -\delta(E+E_\Delta(\vec{k})-M_N)\right)
\;.\end{eqnarray}
{}From eqn. (\ref{rb0}) follows, that the factor in the curly
brackets is equal to $\rho^0_N-1/4\rho^0_\Delta$.
This finding is in contrast to the equations used in
\cite{EWC93,KXS89,XKK93},
where the {\em baryon\/} density has been used instead. The
difference is crucial for the model at higher temperature, where
50--60\% of the baryon density are $\Delta$'s, hence the factor
in the curly brackets is reduced by a factor $\approx 2$ with respect
to the baryon density.

The energy going into the $\Delta$-production is
\begin{equation}\label{odf}
\omega_\Delta(\vec{k})  = E_\Delta(\vec{k}) -M_N
    = \sqrt{\vec{k}^2+M_\Delta^2}-M_N
\;.\end{equation}
Note, that in many applications the above expression is
expand to lowest order in $\vec{k}$,
thus worsening the asymptotic behavior as $\vec{k}\rightarrow\infty$
\cite{EWC93,KXS89,XKK93}.

With this spectral decomposition of the polarization function,
the dispersion integral (\ref{pisd}) can be performed
analytically, giving us the retarded polarization tensor
in quasistatic zero-width approximation as
\begin{equation}
\Pi^R_{q0}(E,\vec{k}) = \vec{k}^2
\frac{8}{9}\left(\frac{f^\pi_{N\Delta}}{m_\pi}\right)^2\,
  \left( \rho^0_N-\frac{1}{4}\rho^0_\Delta\right)
  \frac{\omega_\Delta(\vec{k})}{E^2-\omega^2_\Delta(\vec{k})+
  \mathrm{i}\epsilon^\prime}
\;.\end{equation}
Here $\epsilon^\prime = \mbox{sign}(E)\epsilon$ to obtain the
proper (retarded) boundary conditions in time. The imaginary
part is a $\delta$-function, which gives zero contribution
to the pion propagator.

The spectral function of the pions therefore consists
of isolated poles $\omega_\pm$ obtained as the solution of
\begin{equation}\label{pol}
\omega^2-E^2_\pi(\vec{k}) - \frac{\vec{k}^2\,C\,\omega_\Delta(\vec{k})}{
   \omega^2-E^2_{N\Delta}(\vec{k})}
= 0
\;\end{equation}
with
\begin{eqnarray}
C & = &
  \frac{8}{9}\left(\frac{f^\pi_{N\Delta}}{m_\pi}\right)^2\,
  \left( \rho^0_N-\frac{1}{4}\rho^0_\Delta\right)\nonumber\\
E_{N\Delta}(\vec{k}) & = & \vphantom{\int}\sqrt{
     \omega_\Delta(\vec{k})
     \left( \omega_\Delta(\vec{k}) + g^\prime\,C\right)}
\;.\end{eqnarray}
Form factors depending only on the momentum $\vec{k}$ can
be absorbed into $C$.

For simplicity the momentum arguments are suppressed from now on,
and thus our solutions are
\begin{equation}\label{omf}
\omega^2_\pm = \frac{1}{2}\left(E_{N\Delta}^2+E_\pi^2 \pm
  \sqrt{ \left(E_{N\Delta}^2-E_\pi^2\right)^2 + 4\vec{k}^2 C \omega_\Delta}
  \right)
\;.\end{equation}
The residues of the isolated poles in the propagator are
\begin{equation}
Z_\pm = \frac{1}{2}\left(1 \mp \frac{ E_{N\Delta}^2-E^2_\pi }{
  \sqrt{ \left(E_{N\Delta}^2-E_\pi^2\right)^2 + 4\vec{k}^2 C \omega_\Delta}
  }\right)
\;, \end{equation}
and the retarded propagator itself reads
\begin{eqnarray}
D^{R}_{q0}(E,\vec{k})& =& \frac{E^2-E^2_{N\Delta}}{
  \left(E^2-\omega_+^2+\mathrm{i}\epsilon^\prime\right)
  \left(E^2-\omega_-^2+\mathrm{i}\epsilon^\prime\right)}\nonumber\\
          &=&
  \frac{Z_+}{E^2-\omega_+^2+\mathrm{i}\epsilon^\prime} +
  \frac{Z_-}{E^2-\omega_-^2+\mathrm{i}\epsilon^\prime}
\;.\end{eqnarray}
For the interpretation of these results we consider the free case, where
for small momenta $E_\pi < E_{N\Delta}$, while
at high momenta $E_\pi > E_{N\Delta}$
(see dotted lines in fig. \ref{disp}).

This can be carried over to the interacting case, where for small
momenta the spectral weight $Z_- > Z_+$, hence in this case
$\omega_-$ is the energy of the pion. The other branch in this case
is a $\Delta$-hole excitation.

The roles are exchanged, when the momentum is larger. Then, $\omega_+$
is the energy of the interacting pion with the larger strength $Z_+$,
while $\omega_-$ is associated to the
weaker $\Delta$-hole branch.

The ''crossover'' point is determined by the momentum
$k_\perp$ where the two residues are equal, i.e.
\begin{eqnarray}\label{crx}
E^2_\pi(k_\perp) &=& E^2_{N\Delta}(k_\perp)\nonumber\\
\Leftrightarrow k^2_\perp &=&
\left( \frac{M_\Delta^2-M_N^2-m_\pi^2}{2M_N-g^\prime\,C}
      + M_N\right)^2 - M_\Delta^2
\;.\end{eqnarray}
This value is marked in fig. \ref{disp} by a vertical line.
Hence we can safely set
\begin{eqnarray}\label{omff}
\mbox{pion energy}\;\; \omega_\pi & = & \left\{
   {\array{lll} \omega_- & \mbox{if} & |\vec{k}| < k_\perp \nonumber\\
                \omega_+ & \mbox{if} & |\vec{k}| > k_\perp \endarray}
\right.\\
\mbox{$\Delta$-hole energy}\;\;
\omega_{N\Delta} & = & \left\{
   {\array{lll} \omega_+ & \mbox{if} & |\vec{k}| < k_\perp \\
                \omega_- & \mbox{if} & |\vec{k}| > k_\perp \endarray}
\right.
\;.\end{eqnarray}
The energies $\omega_\pm$ are plotted in figure \ref{disp}
(full lines),
for the example value of 1.69 nuclear density
as function of $|\vec{k}|$. The values of the coupling constants
and masses used in the calculation are given in table 1.

Clearly the notion of a dispersion relation with a discontinuity,
or {\em jump\/}, as implied by the above description,
is somewhat unusual and requires further comment.
We postpone this until
we have discussed a way to include the width of the branches into the
considerations.

In contrast to our derivation, it has been argued
in ref. \cite{EWC93}, that one may introduce the
energies of the quasi-pion and the $\Delta$-hole
excitation as the two combinations
\begin{eqnarray}\label{wr}
\Omega_\pi & = & Z_-\omega_- + Z_+\omega_+ \nonumber\\
\Omega_{N\Delta} & = & (1-Z_-)\omega_- + (1-Z_+)\omega_+
                \; = \; Z_+\omega_- + Z_-\omega_+
\;.\end{eqnarray}
These energies are also plotted
in figure 1 (dash-dotted lines) -- but their choice
is unjustified. They do not correspond to any physical excitation
of the system under consideration.

To prove this we observe, that from the viewpoint of quantum
field theory the renormalization of the
wave function is done with the factors $\sqrt{Z_\pm}$, i.e.,
they are the coefficients of the fields in the dynamical map
\cite{UMT82}. To leading order and for
fixed momentum, the dynamical map for the two interacting fields
can be written explicitly as
\begin{eqnarray}\label{dym}
\psi_+ & =&\sqrt{Z_+}\,{\cal B}_B[E_\pi]\,\xi_\pi
         + \sqrt{Z_-}\,{\cal B}_B[E_{N\Delta}]\,\xi_{N\Delta}
         +\dots \nonumber \\
\psi_- & =&\sqrt{Z_+}\,{\cal B}_B[E_{N\Delta}]\,\xi_{N\Delta}
         - \sqrt{Z_-}\,{\cal B}_B[E_\pi]\,\xi_\pi
         +\dots
\;.\end{eqnarray}
Here, $\xi_\pi$ is the free pion Heisenberg field (thermal
doublet, cf. \cite{Ubook}), transformed with the Bogoliubov
matrix taken at the proper on-shell energy. $\xi_{N\Delta}$
is the operator creating a $\Delta$-hole pair.

Hence, a $Z_\pm$-weighted sum of the eigenmode energies $\omega_\pm$
does not give the energy of a physical state, but up to
factors $\sqrt{Z}$ inverts the
above transformation. This can be seen immediately by calculating,
with the same justification
as can be given for eqn. (\ref{wr}), the averages of the {\em squared\/}
energies. The results then are
\begin{eqnarray}
 Z_-\omega_- ^2+ Z_+\omega_+^2 &=& E_\pi^2\nonumber\\
 Z_+\omega_-^2 + Z_-\omega_+^2 &=& E_{N\Delta}^2
\;,\end{eqnarray}
i.e., the {\em free\/} energies (dotted lines in figure 1).
For all momenta the first choice (\ref{wr})
differs only very slightly from the free energies.

In summary of this comment one can state, that
the $Z_\pm$ weighted energy average
removes the interaction from the system. We therefore feel,
that such an averaging process should be avoided: the resulting
very small modification of the pion dispersion relation
with respect to the free one can be
obtained much easier, than by arguments about polarization functions.
\section{Lorentz spectral function}
The use of a free spectral function for the nucleons is justified
by the great success of the quasiparticle concept in nuclear
physics. It can easily be modified by introducing a density
(and temperature) dependent effective nucleon mass, or collective
effects due to nucleon-hole excitations \cite{h92fock}.
For simplicity however we have kept the bare nucleon, and
therefore use the spectral function
as specified in (\ref{an}).

The $\Delta_{33}$ {\em resonance\/} in contrast has a vacuum width
of 120 MeV. Since this is comparable to the pion mass,
it should not be neglected in any serious computation.
The width of the $\Delta$ spectral function peak also grows
with momentum \cite{R71}, but comparison with scattering data restricts
it to about twice the above value in the momentum range of interest here
\cite{KMO84}.

Even less is known about its medium dependence \cite{KXS89}, hence
without a self-consistent calculation we feel safest to estimate the
influence of a constant $\Delta$-width on the pion spectrum. Such an
approximation then clearly does not have the proper threshold behavior,
but can easily be generalized to such case. For the following, we therefore
use the spectral function
\begin{equation}\label{adw}
{\cal A}_\Delta(E,\vec{p}) = \frac{1}{2\pi}\,
   \frac{\Gamma}{(E-E_\Delta(\vec{p}))^2+(\Gamma/2)^2}
\;.\end{equation}
With the above spectral functions, the actual baryon density instead of
(\ref{rb0}) becomes
\begin{eqnarray}\label{rb}
\rho_b& =& \rho^0_N + \rho_\Delta \nonumber\\
      & =& 4\,\int\!\!\frac{d^3\vec{p}}{(2\pi)^3}\,n_N(E_N(\vec{p}))
          +16\,\int\!\!\frac{d^3\vec{p}}{(2\pi)^3}\,
          \int\!\!dE\, {\cal A}_\Delta(E,\vec{p})\,n_\Delta(E)
\;.\end{eqnarray}
The imaginary part of the polarization function is
\begin{eqnarray}\label{sigf}
\sigma(E,\vec{k})\;&  = &\vec{k}^2
  \frac{16}{9}\left(\frac{f^\pi_{N\Delta}}{m_\pi}\right)^2
  \int\!\!\frac{d^3\vec{p}}{(2\pi)^3}
  \,{\cal A}_\Delta(E_N(\vec{p})+E,\vec{p}+\vec{k})\,\times\nonumber\\
& &    \;\;\;\;\;\;\;\;\;\;\;\;\;\;\;\;\;\;
   \left(\,n_N(E_N(\vec{p}))-\vphantom{\int}
               n_\Delta(E_N(\vec{p})+E)\,\right)\\
&-&\mbox{the same expression with $E\rightarrow -E$}\nonumber
\;,\end{eqnarray}
The complete momentum integral in the second part of (\ref{rb}) can
be done analytically. For the imaginary part of the
polarization tensor in eqn. (\ref{sigf}) however, this only
holds for the angular integration. Using $p=|\vec{p}|$ and
$k=|\vec{k}|$ in the following, we obtain
\begin{eqnarray}\label{sigfc}
\sigma(E,\vec{k})\;  =& \;
  \frac{2\vec{k}^2}{9\pi^3}\left(\frac{f^\pi_{N\Delta}}{m_\pi}\right)^2
  \int\!\!p^2\,dp\,
 \left[
 \left(\,n_N(E_N(p))-\vphantom{\int}n_\Delta(E_N(p)+E)\,\right)
  I(E,k,p)\right.\;& \nonumber\\
 &-\;\left.
 \left(\,n_N(E_N(p))-\vphantom{\int}n_\Delta(E_N(p)-E)\,\right)
  I(-E,k,p)\right]&
\;,\end{eqnarray}
with the function
\begin{eqnarray}
I(E,k,p) &=& \int\!\!d\Omega_p\,{\cal A}_\Delta(E+E_N(p),
\vec{p}+\vec{k})\,\nonumber\\
         &=& \frac{1}{pk}\,\Theta(E_N(p)+E)\,
  \left[ \frac{\Gamma}{2}\,\log\left( x^2+
  \left(\frac{\Gamma}{2}\right)^2 \right)
  \vphantom{\int}  \right.\\
  &&\;\;\;\;\;\;\;\;\;\;\;\;\;\;\;
  \left.\left. -2 (E_N(p)+E)\,\arctan\left(\frac{\Gamma}{2x}\right)
  \right]\right|^{\displaystyle x=w_+}_{\displaystyle x=w_-}\nonumber
\;.\end{eqnarray}
The boundaries to be inserted for $x$ are
\begin{equation}
w_\pm = E_\Delta( p\pm k ) - E_N( p ) - E
\;.\end{equation}
Remaining for the calculation of $\sigma$ is therefore the
$p$-integration, which we perform numerically.
It is followed by the calculation of the dispersion
integral and by the subsequent correction according to eqn.
(\ref{efp}). The resulting spectral function is plotted
as function of temperature and energy for five different
momenta in fig. 2 -- 6.

Clearly, the spectral function has only one peak (as function of energy)
at low momenta (fig.2 and 3) - corresponding to an almost free
pion, with a momentum dependent width. At momenta
in the vicinity of $k_\perp$ according to (\ref{crx}),
this mode is strongly mixed with the $\Delta$-hole
excitation, and two peaks appear in the spectral function
(cf. fig. 4).

At higher momenta, this mixing leads to the crossover of the two
branches, and the diminishing of the peak at lower energies
(fig. 5 and 6). It is worthwhile to note, that the $\Delta$-hole peak
vanishes with increasing temperature at all momenta. Hence
at temperatures above 0.1 GeV, one cannot speak of the
$\Delta$-hole like excitation as a degree of freedom of the system:
it is dissolved in the medium.

Our result is, that the location of the peaks
corresponds quite well to the eigenmodes $\omega_\pm$ of the
quasistatic zero-width approximation, cf. the dots in
fig. \ref{disp}. Hence the latter is useful even at finite temperature.

The same calculations have been performed with a $\Delta$-width of
$2\Gamma$, which therefore allows to study the modifications introduced
by a realistic parameterization of the $\Delta$ spectral function in terms of
scattering data \cite{KMO84}. The results then are quantitatively, but
not qualitatively different from those presented in figs. 2 -- 6. They
only give an even more pronounced smearing of the two spectral peaks
at momenta around $k_\perp$ according to
(\ref{crx}) (cf. fig. 4).
\section{Expansion in lowest order $\Gamma$}
The calculation of dispersion integrals with sufficient precision
appears to be impractical for many purpose. Hence in the following
we study a modification of the quasistatic zero-width
approximation, which takes into account the finite width
effects in a controlled fashion.

To this end, we start from eqn. (\ref{sigf}) and
perform a quasistatic approximation as before
by neglecting the $\vec{p}$-dependence only in the spectral function of the
$\Delta$'s. Care has to be taken, however, because of the rather
awkward factor from the second distribution function. Hence we assume,
that ${\cal A}_\Delta$ is still quite peaked around the bare
energy value. Then we have approximately
\begin{equation}\label{sig1}
\sigma_{q}(E,\vec{k})  =\vec{k}^2
  \frac{16}{9}\left(\frac{f^\pi_{N\Delta}}{m_\pi}\right)^2\,
\left(\rho^0_N-\frac{1}{4}\rho^0_\Delta\right)
 \left( {\cal A}_\Delta(M_N+E,\vec{k})\,
         -{\cal A}_\Delta(M_N-E,\vec{k})\right)
\;.\end{equation}
Note, that still $\rho_\Delta^0$ appears in the coupling factor,
rather than $\rho_\Delta$ as defined in (\ref{rb}).

Now we follow the approach from \cite{SRS90} and
perform an asymptotic expansion of the $\Delta$ spectral
function
\begin{equation}\label{adw2}
{\cal A}_\Delta(E,\vec{p}) = \delta(E-E_\Delta(\vec{p}))- \frac{1}{2\pi}\,
   \frac{\partial}{\partial E}\,\frac{{\cal P}\;\Gamma}{E-E_\Delta(\vec{p})}
\;.\end{equation}
Here, ${\cal P}$ denotes the principal value in case one integrates
over the energy pole. Note, that this equation also holds in case
$\Gamma$ acquires an $E$, $\vec{k}$, $T$-dependence.
Using eqn. (\ref{sig1}), this yields
\begin{eqnarray}\label{sig2}
\sigma_{q1}(E,\vec{k}) & = \displaystyle\vec{k}^2
  \frac{16}{9}\left(\frac{f^\pi_{N\Delta}}{m_\pi}\right)^2\,
&\left(\rho^0_N-\frac{1}{4}\rho^0_\Delta\right)
 \left( \vphantom{\int}\delta(E-E_\Delta(\vec{k})+M_N)\right.
\nonumber\\
       &-&\left.\delta(E-E_\Delta(\vec{k})-M_N)
       +\frac{\Gamma\,E}{\pi}\,
       \frac{\omega_\Delta(\vec{k})}{
   \left(E^2-\omega_\Delta^2(\vec{k})\right)^2}
   \right)
\;.\end{eqnarray}
Only the $\delta$-functions contribute to the dispersion integral
(\ref{pisd}), hence to first order in $\Gamma$ the
polarization function is
\begin{eqnarray}
\label{ga1}
\Pi^R_{q1}(E,\vec{k}) & =&\vec{k}^2C\,
   \frac{\omega_\Delta}{E^2-\omega_\Delta^2}
   -\vec{k}^2C \,\mathrm{i} \,\frac{\omega_\Delta E\,\Gamma}{
                            \left(E^2-\omega_\Delta^2\right)^2}\\
\label{ga2}
& \approx& \vec{k}^2C\,\frac{\omega_\Delta}{
                         (E+\mathrm{i}\Gamma/2)^2-\omega_\Delta^2}
\; \end{eqnarray}
(before the $g^\prime$-correction according to (\ref{efp})).
The pion propagator obtained with this approximation has two poles
in the proper complex energy half plane, and is an analytical function
in the other half plane.

Another approximation used frequently is obtained by simply shifting
$\omega_\Delta$ by $\mathrm{i}\Gamma/2$ as in refs. \cite{KLQ93,XKK93}:
\begin{equation}\label{ko}
\vec{k}^2C\,\frac{\omega_\Delta-\mathrm{i}\Gamma/2}{
                        E^2-(\omega_\Delta-\mathrm{i}\Gamma/2)^2}
\;.\end{equation}
However, in contrast to our derivation above, the
pion propagator obtained with (\ref{ko}) has not the proper
structure in the complex energy plane (see the comment below and
fig. \ref{asy035}).

After the correction (\ref{efp}) is performed, the correct pionic
spectral function obtained with the expression (\ref{ga2}) is,
to first order in $\Gamma$,
\begin{equation}\label{asys}
{\cal A}_\pi(E,\vec{k}) =
\frac{\vec{k}^2C}{\pi}\,\frac{\Gamma\,E\,\omega_\Delta\,\Theta(E)}{
   (E^2-\omega^{\prime\,2}_+)^2\,
   (E^2-\omega^{\prime\,2}_-)^2 +  \Gamma^2\,E^2\,(E^2-E^2_\pi)^2}
\;.\end{equation}
Here, to preserve unitarity, the energies in the denominator have
been obtained in slight modification of (\ref{omf}) as
\begin{equation}\label{omf2}
\omega^{\prime\, 2}_\pm = \frac{1}{2}\left(E_{N\Delta}^2
 +(\Gamma/2)^2+E_\pi^2 \pm
  \sqrt{ \left(E_{N\Delta}^2+(\Gamma/2)^2
-E_\pi^2\right)^2 + 4\vec{k}^2 C \omega_\Delta}
  \right)
\;.\end{equation}
In fig. \ref{asy035}, we compare the results for the three
forms of the asymptotic expansion, obtained with eqns.
(\ref{ga1}) (dash-dotted line), (\ref{ga2}) (full line)
and (\ref{ko}) (dotted line) to the full calculation from the previous
section (dashed line). At zero temperature near
the point of maximal mixing $|\vec{k}|\approx k_\perp$
the first form (\ref{ga1}) is rather crude and hence will be
avoided henceforth. The approximation (\ref{ga2}) however
corresponds quite well to the full calculation - while the
third form (\ref{ko}) contributes unphysical strength at low
energies, at least for the energy independent $\Gamma$ used here.

While an explicitly energy dependent $\Gamma$ might cure this
(eqn. (\ref{adw2}) still holds in this case), for our discussion
of an energy-independent $\Gamma$
we thus consider eqn. (\ref{ga2}) the best choice
for an approximate treatment of the $\Delta$-width.
For zero temperature, this agreement prevails at all momenta
up to twice nuclear density, see the upper panel in fig.
\ref{den035}.

However, the picture changes substantially when
looking at finite temperatures - and this was the main goal
of the present work. We find, that at finite temperature the asymptotic
expansion severely underestimates the smearing of the spectrum.
This is shown in the small inserts to fig. \ref{cgfit035} and the lower panel
of fig. \ref{den035}: the discrepancy increases with temperature,
but is almost independent of density in the range considered here.

The deviation is connected to the smearing of the
Fermi surface, i.e. to the quasistatic approximation leading from
(\ref{sigf}) to (\ref{sig1}) rather than to the asymptotic
equation (\ref{adw2}). It can be corrected by fitting
an ''effective'' $\Gamma^\prime(\vec{k},T)$ to the full calculation.

In the main body of fig. \ref{cgfit035} and the lower panel of fig.
\ref{den035} we show the results of such a fit.
Both figures contain a comparison of the full calculation from the previous
section (thick lines) to the asymptotic expansion (thin lines).
The asymptotic expansion has been taken from eqn. (\ref{ga2}),
but with a modified $\Gamma^\prime(\vec{k},T)$ -- whereas the small
inserts to these figures are calculated with the $\Gamma$ from table 1.

The effective $\Gamma^\prime(\vec{k},T)$ is plotted in fig. \ref{gfit}.
Of course, such a momentum and temperature dependence also arises
from a medium-dependent $\Delta$ spectral function or its momentum
dependent parameterization \cite{KMO84}. However,
we find that even when starting from eqn. (\ref{adw})
with a constant $\Gamma$, the pion propagation
feels an effectively medium dependent $\Gamma^\prime$.
The changes here are of similar magnitude as one would expect
from a self-consistent calculation of the $\Delta$-width \cite{KXS89}.
The additional width is additive, i.e., if the bare width of the
$\Delta$ is doubled, the effective width according to our fit does
not increase by a factor of two.

It is now clear, how to construct a simple approximation to the pionic
spectral function in HCNM. This can be used in the calculation
of observable quantities. However, for the practical purpose of
simulations as well as for a qualitative understanding, a
particle-like picture would be more appealing. We therefore attempt to
define an interacting pion propagating in HCNM.

Such description is far from trivial, since one
cannot simply look for poles of the pion propagator in the complex
plane: they are on an unphysical Riemann sheet.
Instead, one has to define the particle-like excitations
(which are {\em not\/} quasi-particles) on the
real energy axis. To this end, we follow a procedure recently
outlined for non-relativistic systems \cite{CU93},
and write the inverse retarded
propagator for the pions with real energy $k_0$ as
\begin{equation}
k_0^2 - E^2_\pi(\vec{k}) - \Pi^R(k_0,\vec{k})
  = \left( k_0 -( \varepsilon_k-\mathrm{i}\gamma_k)\right)
    \left( k_0 +( \varepsilon_k+\mathrm{i}\gamma_k)\right)
\;.\end{equation}
To lowest order, the on-shell energy $\varepsilon$ and its imaginary
part $\gamma$ are then given as the solution of the equation
\begin{equation}\label{onsh}
\left( \varepsilon_k-\mathrm{i}\gamma_k\right)^2
  - E^2_\pi(\vec{k}) - \Pi^R(\varepsilon_k,\vec{k}) =0
\;.\end{equation}
Using the asymptotic expansion (\ref{ga2}) after the correction
according to (\ref{efp}), the solution of this equation
corresponds to solving two coupled (non-linear) polynomial
equations. Although this is easily done numerically, one cannot
guarantee the existence of a unique solution. Indeed, for a broad
range of momenta {\em two\/} solutions exist. They are shown as
full thin lines in fig. \ref{disp}, the shaded areas correspond
to the regions $\varepsilon_k+\gamma_k > E >
\varepsilon_k-\gamma_k$.

The fact, that the real part of the energies $\varepsilon_k$
corresponds quite well to the peaks of the full calculation from
section 4 proves the validity of the approximation (\ref{onsh}).
Moreover, as can be seen from fig. \ref{onshf}, the imaginary
part of the on-shell energy is comparable to the half-width of
the spectral function peaks. Thus we conclude, that the
definition (\ref{onsh}) is an easy way to determine the
properties of pionic modes in HCNM.

It remains to clarify, which of the two branches to chose for a
particle-like pion propagation. A natural choice is, to take as
''the pion'' that solution of eqn. (\ref{onsh}), which has the
smallest imaginary part of the energy. This choice is unique,
since one of the solutions has an imaginary part increasing with
momentum, while the other decreases with $|\vec{k}|$.  Thus,
there exists a certain momentum, where one has to switch from one
branch to the other with increasing or decreasing momentum.  As
one can find analytically from the above equations, this
crossover point is given by the momentum value, where the two
residues in the quasistatic zero-width approximation are equal,
i.e., by the $k_\perp$ defined in eqn. (\ref{crx}) (cf. also fig.
\ref{disp} and the vertical lines in fig. \ref{onshf}).

The  real and imaginary part of the pion dispersion relation
obtained with this description are shown in fig. \ref{onshf}, for
several temperatures at 1.69 nuclear density. To keep the picture
as clean as possible, we have performed these calculations with
the constant $\Gamma$ from table 1. In fig. \ref{onshdenf} we
show the same quantities at three different densities.

In short, the peculiarity of our definition is a {\em jump\/} in the dispersion
relation of the interacting pion (To guide the eye, the curves in
the figures have been drawn continuously.).
Such a discontinuity in the dispersion relation can be seen as a
potential well in momentum space: If the pion momentum gets
smaller than the critical value of $k_\perp$, the energy drops by
a certain amount. This drop diminishes with increasing
temperature (see fig. \ref{onshf}) and decreasing density (see
fig.  \ref{onshdenf}). The energy released in this process is
distributed into the medium by the coupling to the broad
$\Delta$-hole excitation (cf. fig. 4) .

The imaginary part of the quasi-pion energy has
its maximum at the boundary of the well, i.e.,
at momentum $k_\perp$. As can be derived, this maximum is
$\mathrm{max}(\gamma_k)=\Gamma/2$.
The quasi-pion therefore moves through
the nuclear medium with a complex velocity, similar to a photon
crossing a semi-transparent medium.
\section{Conclusions}
In this work, the formalism of Thermo Field Dynamics was applied
to the $\Delta$-hole model at finite temperature. The model
was analyzed starting from a seemingly simple quasistatic zero
$\Delta$-width approximation, which can be treated analytically.
However, as an important difference to other treatments
of the model, a diminishing of the coupling was found at higher
temperature (cf. eqn. (\ref{sig0}) and its discussion,
and compare to refs. \cite{EWC93,XKK93}).

Furthermore we could show, that the residue-weighted energy average
of the eigenmodes in this simple approximation does not give a proper
description of pion propagation: one could as well use the free
pion in the system.

We then went to a more elaborate treatment, taking into account
a constant width of the $\Delta_{33}$ resonance. The numerical
calculation of the pionic spectral function at finite temperature
was performed to establish a reference point for simplifications.
Of these the most promising was found by performing an asymptotic
expansion of the $\Delta$ spectral function, leading to
a polarization function (\ref{ga2}) that is a straightforward
generalization of the zero-width approximation. This expansion
can also be applied to more sophisticated parametrizations
of the $\Delta$ spectral function, where the width is
$E$, $\vec{k}$, $T$ - dependent.

Our approximation agrees quite well with the full reference
calculation, thus supplying a simple expression for the full
pion propagator in hot, compressed nuclear matter (HCNM).
Another approximation however was found to produce
artificial spectral strength at low energies (cf. eqn. (\ref{ko}),
fig. \ref{asy035} and ref. \cite{KLQ93}).

Since the Bose
equilibrium distribution function is strongly peaked at small
energies, this strength leads to
a large increase of the effective (near-) equilibrium occupation number.
While such spectral contribution at small energies might
therefore explain the low momentum pion enhancement
in relativistic heavy-ion collisions \cite{XKK93,O88}, it is clearly
not a consequence of a rigorous treatment of the $\Delta$-hole
model.

The successful description of phenomena like lepton pair
creation in heavy-ion collisions requires to use the full pion
propagator according to (\ref{dbk}), with one or the
other spectral function as discussed in this work \cite{HFN93,KLQ93}.
However, for cascade-like simulation calculations
as well as for other practical purpose, a description
of ''the pion'' in terms of spectral functions and matrix valued
propagators might be impractical.

We therefore gave a simplified description, i.e. defined
particle-like excitations with a definite (complex) energy.
While the real part of this dispersion relation coincides
very well with the simple quasistatic zero-width approximation,
our model also gives the spectral width of the pion-like
excitation. The real part of the
dispersion relation obtained in this way includes a discontinuity,
i.e., a potential well in momentum space -- and the absorptive
part of the dispersion relation is maximal at the boundary of the
well. Hence, this simplified
description of pionic modes in HCNM corresponds to a kind of
''pion-optical'' picture.

One of the implications of this picture is, that the (near-)equilibrium
occupation number of pions is enhanced for pion momenta just inside
the potential well, while it is decreased for momenta just above
the step (cf. eqn. (\ref{crx})). Further implications, e.g.
on pion spectra obtained from nuclear collisions, will have to be
investigated.

Furthermore one has to add, that a fully self-consistent
treatment of the medium- and temperature dependent $\Delta$-width
is desirable. Without such considerations, a simple
parameterization of our results as function of temperature and
density is meaningless.  However, since we derived a
quite large effect of temperature on the pion width, small
changes from such calculation will not modify our conclusions. In
essence, one has to solve the Fock problem for pions, nucleons
and $\Delta$'s - and for this, solutions exist that can
be supplemented by the pionic spectral functions we derived
\cite{h92fock}.

We did not discuss negative energy states and renormalization
problems in the present paper. It is nevertheless clear that they
play an important role for model consistency: Although the
$\Delta$-$\bar{N}$ continuum begins at much higher energies than
discussed here, the polarization function has to fulfil the
dispersion relation (\ref{pisd}). Hence, even at low energies
mesonic spectral functions are influenced by the continuum
\cite{dfh93}.
\begin{ack}
One of the authors (P.H.) wishes to express his thanks to
Gy. Wolf, M.Herrmann and B.Friman for numerous discussions and
valuable comments.
\end{ack}

\begin{figure}
\vspace*{165.4mm}
\includegraphics{dispef.ps}
\caption{Pionic dispersion relation at $\rho_b$ = 1.69 nuclear density.}
\label{disp}
\small
Full thick lines: Quasistatic zero-width approximation,
 eqn. (\ref{pol})\\
Dotted lines: Free energies $E_\pi$ and $E_{N\Delta}$,
 vertical lines $k_\perp$ from eqn. (\ref{crx})\\
Dash-dotted lines: $Z_\pm$ weighted energies, eqn. (\ref{wr})\\
Dots: Location of the peaks in the spectral function of the full
calculation, (\ref{sigfc}).\\
Full thin lines and hatched area: Asymptotic expansion with on-shell
  definition, eqn. (\ref{onsh}). Hatched is the region within
$\varepsilon_k\pm\gamma_k$.
\normalsize
\end{figure}
\begin{figure}
\vspace*{165.4mm}
\includegraphics{a5005.ps}
\caption{Pionic spectral function at $|\vec{k}|$ = 50 MeV,
$\rho_b$ = 1.69 nuclear density.}
\end{figure}
\begin{figure}
\vspace*{165.4mm}
\includegraphics{a5020.ps}
\caption{Pionic spectral function at $|\vec{k}|$ = 200 MeV.}
\end{figure}
\begin{figure}
\vspace*{165.4mm}
\includegraphics{a5035.ps}
\caption{Pionic spectral function at $|\vec{k}|$ = 350 MeV.}
\end{figure}
\begin{figure}
\vspace*{165.4mm}
\includegraphics{a5050.ps}
\caption{Pionic spectral function at $|\vec{k}|$ = 500 MeV.}
\end{figure}
\begin{figure}
\vspace*{165.4mm}
\includegraphics{a5065.ps}
\caption{Pionic spectral function at $|\vec{k}|$ = 650 MeV.}
\end{figure}
\begin{figure}
\vspace*{165.4mm}
\includegraphics{asy035.ps}
\caption{Pionic spectral function at T=0 and
$\rho_b$ = 1.69 nuclear density}
\label{asy035}
\small
Full line: Asymptotic expansion, eqn. (\ref{ga2})\\
Dashed line: Full calculation from section 4\\
Dotted line:  eqn. (\ref{ko}),
Dash-dotted line: eqn. (\ref{ga1}).
\normalsize
\end{figure}
\begin{figure}
\vspace*{165.4mm}
\includegraphics{cgfit035.ps}
\caption{Pionic spectral function at T=0.1 GeV
and $\rho_b$= 1.69 nuclear density}
\label{cgfit035}
\small
Thick lines: Full calculation from section 4\\
Thin lines: Asymptotic expansion, eqn. (\ref{ga2})\\[1mm]
Temperatures 50 MeV (full lines), 100 MeV (dashed) and 150 MeV
(dash-dotted)\\[1mm]
Main panel calculated with $\Gamma^\prime(\vec{k},T)$ from fig.
   \ref{gfit}.\\
Insert calculated with $\Gamma$=0.12 GeV.
\normalsize
\end{figure}
\begin{figure}
\vspace*{165.4mm}
\includegraphics{den035.ps}
\caption{Pionic spectral function at $T$=0 and $T$=0.1 GeV}
\label{den035}
\small
Thick lines: Full calculation from section 4\\
Thin lines: Asymptotic expansion, eqn. (\ref{ga2})\\[1mm]
$\rho_b$=  1.69 (full lines), 0.92 (dashed)
and 0.47 (dash-dotted) nuclear density.\\
Lower main panel calculated with $\Gamma^\prime(\vec{k},T)$ from fig.
   \ref{gfit}.\\
Top panel and insert calculated with $\Gamma$=0.12 GeV.
\normalsize
\end{figure}
\begin{figure}
\vspace*{165.4mm}
\includegraphics{gfit.ps}
\caption{Effective thermal $\Delta$-width
$\Gamma^\prime(\vec{k},T)$ in GeV
at $\rho_b$ = 1.69 nuclear density.}
\label{gfit}
\small
Both panels show the same data, contour spacing in the lower
panel is 10 MeV.
\normalsize
\end{figure}
\begin{figure}
\vspace*{165.4mm}
\includegraphics{onsh.ps}
\caption{Pion dispersion relation at
$\rho_b$ = 1.69 nuclear density.}
\label{onshf}
\small
Upper panel real part $\varepsilon_k$, lower panel imaginary part $\gamma_k$
of the effective pion energy from (\ref{onsh}).\\
Temperatures 1 MeV (full lines), 50 MeV (dashed),
100 MeV (dash-dotted), 150 MeV (dash-double-dotted)
and 200 MeV (dotted).\\
Full thin line: free pion, vertical lines $k_\perp$ from
eqn. (\ref{crx}).
\normalsize
\end{figure}
\begin{figure}
\vspace*{165.4mm}
\includegraphics{onshden.ps}
\label{onshdenf}
\small
\caption{Pion dispersion relation at $T$=0 and $T$=0.1 GeV.}
Upper panel real part $\varepsilon_k$, lower panel imaginary part $\gamma_k$
of the effective pion energy from (\ref{onsh}).\\
Temperatures 1 MeV (thin lines) and 100 MeV (thick lines).\\[1mm]
Baryon densities $\rho_b$=  1.69 (full lines), 0.92 (dashed)
and 0.47 (dash-dotted) nuclear density.\\
Dotted line: free pion
\normalsize
\end{figure}
\end{document}